\documentclass[12pt,a4paper]{article}
\usepackage{amsmath}
\usepackage{slashed}
\usepackage{amssymb}
\usepackage[usenames, dvipsnames]{color}
\usepackage{mathtools}
\usepackage{verbatim}
\usepackage{amsfonts}
\usepackage{mathrsfs}
\usepackage{graphicx}
\usepackage[font=small]{caption}
\usepackage[caption=false]{subfig}
\usepackage{enumitem}
\usepackage[normalem]{ulem}
\usepackage[percent]{overpic}
\usepackage{bm}
\usepackage{bbm}
\usepackage{rotating}
\usepackage{hyperref}
\usepackage{cancel}
\usepackage{verbatim}
\usepackage{graphicx}
\usepackage{tensor}
\usepackage{wrapfig}
\usepackage{booktabs}
\usepackage{natbib}
\usepackage{setspace}
\usepackage{relsize}
\usepackage{authblk}

\usepackage[hang,flushmargin]{footmisc}
\AtBeginEnvironment{footnote}{\footnotesize}
\usepackage{geometry}
\usepackage{braket}

\newcommand{\GG}[1]{}

\begin{document}
\title{From Humean Laws to a Neo-Kantian Spacetime:\linebreak
A Dynamics-First View of Topology}
\author{\vspace{-0.25cm} Daniel Grimmer}\footnotetext{\ \ Email: daniel.grimmer@philosophy.ox.ac.uk\\
\ \ Orchid ID: 0000-0002-8449-3775}
\affil{\small \vspace{-0.25cm} Faculty of Philosophy, University of Oxford, Oxford, OX2 6GG United Kingdom\\
Reuben College, University of Oxford, Oxford, OX2 6HW United Kingdom}
\date{}

{\singlespacing \maketitle}
\vspace{-1.5cm}
\begin{abstract}
Do the spacetime manifolds which feature in our best scientific theories reflect anything metaphysically weighty in the world (e.g., any fundamental substances or relations)? Should we extend our notions of space and time beyond the epistemological roles they play in helping us codify the dynamical behavior of matter? Kant famously answered ``No'' to both of these questions, contra Newton and Leibniz. This paper introduces novel technical and philosophical support for such a \mbox{(Neo-)Kantian} perspective on the metaphysics of space and time. To begin, I will make an explicit analogy between broadly Humean views of laws (e.g., Lewis, Demarest, etc.) and dynamics-first views of geometry (e.g., Brown). I will then continue this line of analogous views beyond the metaphysics of laws debate and the dynamical vs geometric spacetime debate by extending it into the context of spacetime topology. Namely, I will put forward a dynamics-first view of topology (in answer to Norton's problem of pre-geometry). 

This dynamics-first view of topology is supported by some powerful new techniques for topological redescription which I have recently developed in \cite{IntroISE} and \cite{ISEEquiv}, namely the ISE Method. These new techniques allow us to remove and replace the topological underpinnings of our spacetime theories just as easily as we can switch between different coordinate systems. For instance, a theory set on a M\"{o}bius strip might be redescribed as being set on the Euclidean plane and vice versa. Indeed, as \cite{ISEEquiv} has proved, the ISE Method gives us access to effectively \textit{every possible} spacetime framing of a given theory's kinematical and dynamical content. Given this overabundance of candidate spacetime framings, it is then conceivable that one can pick out a theory's topological structure via something analogous to a Best Systems Analysis.
\end{abstract}

\section{Introduction}



What, if anything, can help us explain the dynamical behavior of matter?\footnote{This paper considers only metaphysically supported explanations, i.e., explanations which require that we grant some metaphysical weight to their explanans. The central issue in this paper is whether or not one should grant any metaphysical weight to various theoretical structures (e.g., laws, coordinates, geometry, topology, etc.) in hopes of giving a metaphysically supported explanation of the dynamical behavior of matter. Dynamics-second views do this whereas dynamics-first views do not. Instead, dynamics-first views offer us a metaphysically supported explanation that is rooted in dynamics, e.g., the laws of nature are what they are ultimately because of the dynamical behavior of matter, not vice versa. On this view, the laws might still help us ``explain'' dynamics but not in a metaphysically supported way. For instance, Newton's law of gravity provides us with a unified account of the motions of apples and planets without giving us a metaphysical picture of how gravitation works. This would be an explanation by unification, see~\cite{ExplainUnify}.
\label{FnMetaExplain}} Allow me to first demonstrate several candidate explanans by considering four example questions about dynamics (whose answers I will later challenge). Q1: Why did that apple fall? A1: Because of the laws of gravity. Q2: Why does Foucault's pendulum rotate? A2: Because the Earth's lines of latitude and longitude do not form an inertial coordinate system. Q3: Why do all kinds of clocks exhibit time dilation no matter what material they are made of? A3: Because the geometric structure of spacetime is Lorentzian. Q4: On a spacetime manifold with spherical topology, why do our measuring rods always measure the average spatial curvature to be positive no matter the metric field? A4: Because the topological structure of this spacetime manifold has an Euler characteristic of $\chi=2$ (this then has implications for curvature by the Gauss-Bonnet theorem). Thus, we can identify four candidate explanans for dynamics: laws, coordinates, geometry, and topology.

But how have we come to know anything about these candidate explanans? The story goes roughly as follows: Scientists have carefully studied the observable dynamics of matter (e.g., the motion of certain needles in a laboratory). Then, after a significant amount of work, they have codified these observations into our current best scientific theories. Notably, these theories will typically refer to matter and dynamics beyond those which were directly observed (e.g., they may refer to electrons, currents, magnetic fields, etc.). Moreover, it is common for these theories to feature additional structures which do not refer to either matter or its dynamical behavior. Namely, they might feature laws of nature and/or special coordinate systems. They might also make use of some fixed geometric and topological structures (e.g., the fixed Lorentzian metric, $\eta_\text{ab}$, and spacetime manifold, $\mathcal{M}$, appearing in special relativity). Notice that these four additional theoretical structures are exactly our four candidate explanans.

\begin{figure}[p!]
\begin{center}
\includegraphics[width=0.95\textwidth]{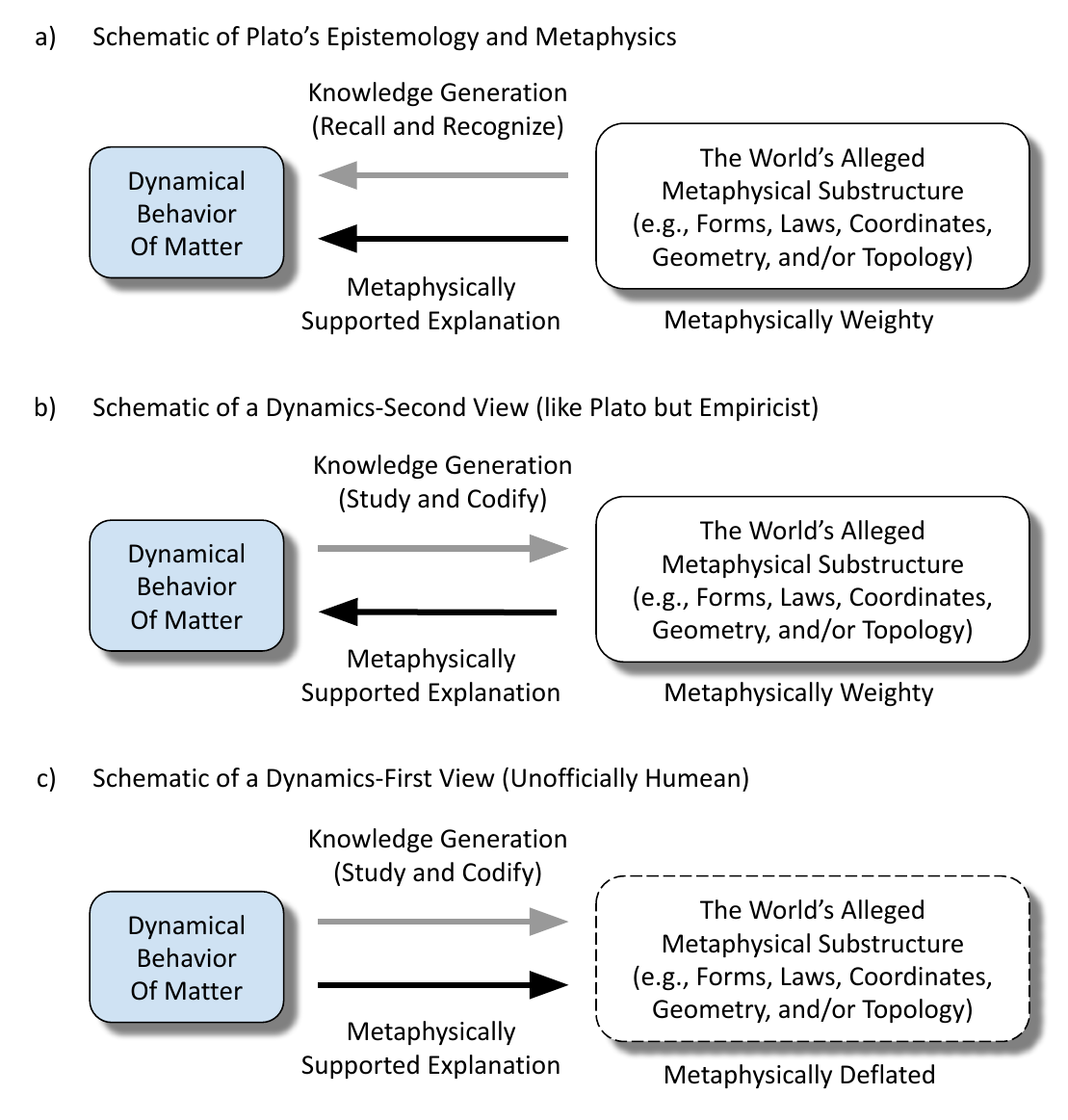}
\caption{This figure shows a schematic of Plato's epistemology and metaphysics (a), as well as schematics of a dynamics-second view (b) and a dynamics-first view (c). For each kind of view, this figure shows the direction of the arrow of knowledge generation as well as the direction of the arrow of metaphysically supported explanation. Plato's view has the Forms as the ultimate source of both knowledge and metaphysical explanation. Dynamics-second views continue with Plato's style of metaphysics while rejecting his epistemology in favor of empiricism. By contrast, for dynamics-first views additionally reverse the arrow of explanation. As I discuss in the main text, this move is of an unofficially Humean character and ultimately allows for a sparser metaphysics.}\label{FigDynamicsSecondFirst}
\end{center}
\end{figure}

Hence, we know where our knowledge of these four candidate explanans comes from. Regardless of whether these theoretical structures can or cannot help us explain dynamics, we can all agree that we have learned about them by studying and codifying the dynamical behavior of matter. Let us call this the \textit{arrow of knowledge generation}. It is after this epistemological point of agreement, however, that some metaphysical disagreements may arise regarding the direction of the arrow of explanation. In each of the four contexts (e.g., laws, coordinates, geometry, and topology) one might take the arrow of explanation to run antiparallel to the arrow of knowledge, see Fig.~\ref{FigDynamicsSecondFirst}b). Namely, one might follow A1-A4 in appealing to these theoretical structures in order to explain dynamics. Alternatively, however, one might reverse the arrow of explanation so that it runs parallel to the arrow of knowledge, see Fig.~\ref{FigDynamicsSecondFirst}c).\footnote{Some have tried to find a middle ground between the dynamics-first and dynamics-second approaches by claiming that the arrow of explanation is bi-directional. For instance, \cite{Acuna2016} claims that Minkowski spacetime structure and Lorentz invariant dynamics are two sides of a single coin.} That is, one might explain these theoretical structures in terms of dynamics, e.g., the laws of nature are what they are ultimately because of the dynamical behavior of matter, not vice versa.

In the next section, I will unite the pre-existing debates about the metaphysics of laws, coordinates, and geometry into a single framework, which I will call the dynamics-first vs dynamics-second debates. While these three debates are already well-established individually, a detailed discussion of their similarities is conspicuously missing from the existing literature. However, the following section does more than just provide a novel perspective on the relevant literature. My ultimate goal is to extend these debates beyond the contexts of laws, coordinates, and geometry, into the context of spacetime topology. In fact, I will put forward a dynamics-first view of topology in Sec.~\ref{SecISEFTop} (with the help of Secs.~\ref{SecTheoryLevel} and~\ref{SecRedescription}). This new view is intended to be closely analogous to the dynamics-first views of laws, coordinates, and geometry discussed below. Being so analogous, these four dynamics-first views share in their motivations and lend each other mutual support. Since my argument for this dynamics-first view of topology relies upon this analogy, some substantive philosophical work is first needed in unifying these dynamics-first vs dynamics-second debates.

\section{Four Dynamics-First vs Dynamics-Second Debates}\label{Sec2v1}

According to what I will call \textit{dynamics-second views} (of laws, coordinates, geometry, topology, etc.), our theoretical investigations of matter's dynamics have revealed to us the metaphysical substructure of the world (e.g., its nomological, geometric, topological substructures, etc.). Moreover, these revealed substructures are taken to have enough metaphysical weight that we can then turn around and use them to help give a metaphysically supported explanation of the very dynamics which we began investigating. See Fig.~\ref{FigDynamicsSecondFirst}b). For instance, the governing view of laws is a dynamics-second view:\footnote{More generally, dynamics-second views of laws are roughly what \cite{Ott2022} calls the ``top-down'' views in the metaphysics of laws debate: e.g., Cartesian occasionalism, see~\citeauthor{Ott2009} (\citeyear[Ch. 9]{Ott2009}); primitivism, see~\citeauthor{Ott2022} (\citeyear[Ch. 4]{Ott2022}); and the Dretske–Tooley–Armstrong conception of laws as relations between universals, see~\cite{ArmstrongNomNec}.} It claims that by investigating some dynamical phenomena we can discover the world's nomological substructures (i.e., the laws of nature) which in turn \textit{govern} those very phenomena. 

Relatedly, the following story provides the outline for a dynamics-second view of geometry:\footnote{The dynamics-second views of geometry are on the geometrical side of the geometrical vs dynamical spacetime debate. This largely orthodox view is implicit in \citeauthor{Friedman1983}'s textbook (\citeyear[Ch 6 Sec 4]{Friedman1983}). It can be seen more explicitly in \cite{MaudlinTim2012Pop:,JANSSEN200926,Norton2008,DoratoMauro2007RTbS}.} By studying the inertial behavior of measuring rods and clocks we can discover the world's geometric substructure (e.g., a Lorentzian metric, $\eta_\text{ab}$). In turn, this geometric substructure then serves to constrain the symmetries of all dynamics and thereby helps us to explain the very phenomena under consideration. Finally, a dynamics-second view of topology could be built around the following story: By studying and codifying the dynamical behavior of matter we can discover the world's topological substructure (e.g., a spacetime manifold, $\mathcal{M}$). In turn, this topological substructure then gives the dynamics a spatiotemporal arena to happen within and thereby helps us explain this dynamics.

This dynamics-second explanatory strategy can be traced all the way back to Plato. For Plato, the world's metaphysical substructures were the Platonic Forms and these were taken to be the ultimate explanans for the physical world, see~\cite{Plato1997Cwer}. Hence, like the dynamics-second views, Plato's arrow of explanation also points toward dynamics from some metaphysically weighty substructures. The key difference, however, is that for Plato the arrow of knowledge also points towards dynamics from these substructures. Namely, according to Plato, we don't learn about the Forms by studying the physical world; Instead, we learn about the physical world by recognizing it as an imitation of the Forms which we can recall from having previously interacted with them directly, see~\cite{PlatoMeno}. Compare  Fig.~\ref{FigDynamicsSecondFirst}a) and Fig.~\ref{FigDynamicsSecondFirst}b). Note that dynamics-second views accept Plato's style of metaphysics while rejecting his epistemology in favor of empiricism.

Some philosophers (myself included) find the metaphysical substructures invoked by such dynamics-second explanations to be excessively heavy. A metaphysically lighter way forward is to reverse the arrow of explanation, putting dynamics first, so to speak. See Fig.~\ref{FigDynamicsSecondFirst}c). On such \textit{dynamics-first views} (of laws, coordinates, geometry, topology, etc.) one still codifies the dynamical behavior of matter into theories that can feature nomological, geometric, and/or topological structures. At this point, however, one does not grant the relevant theoretical structures any metaphysical weight of their own, nor does one attempt to use them as explanans for dynamics. Instead, one limits these additional theoretical structures to the epistemic roles which they play in helping us to codify dynamics. To see some examples of such dynamics-first views, let us first look to the metaphysics of laws debate.\footnote{What I am calling the dynamics-first vs dynamics-second debate about laws is, of course, a well-established debate within the metaphysics of laws literature. For an overview of this debate see the SEP article by \cite{sep-laws-of-nature} among many others: \cite{Demarest2017,Lewis1973,LewisDavid1983Nwfa,Lewis1999, BhogalHarjit2020Halo,Hall2015,Ott2022,ArmstrongNomNec,BirdTextbook}, and \cite{BusinessOfLaws}.}

\subsection{Dynamics-First Views of Laws}\label{SecDFLaws}

Humeanism about laws is a good example of a dynamics-first view of laws. It claims (among other things) that the laws which appear in our best theories reflect nothing metaphysically weighty in the world beyond them being a particularly nice way of codifying the dynamical behavior of matter.\footnote{As I will discuss at the end of Sec.~\ref{SecDFLaws}, different dynamics-first views of laws are allowed to vary regarding what the term ``particularly nice'' means exactly.} Reversing the arrow of explanation in this way allows us to metaphysically deflate the laws and thereby avoid the following tricky metaphysical questions. How is matter supposed to ``know'' what the laws are so that it can behave accordingly? What gives the laws the ``nomic voltage'' required to govern matter? How exactly do the laws govern matter? By pushing it around, so to speak? If so, then why can't the matter push back on the laws? Such questions are ubiquitous throughout the metaphysics of laws debate.

\begin{figure}[t!]
\begin{center}
\includegraphics[width=0.9\textwidth]{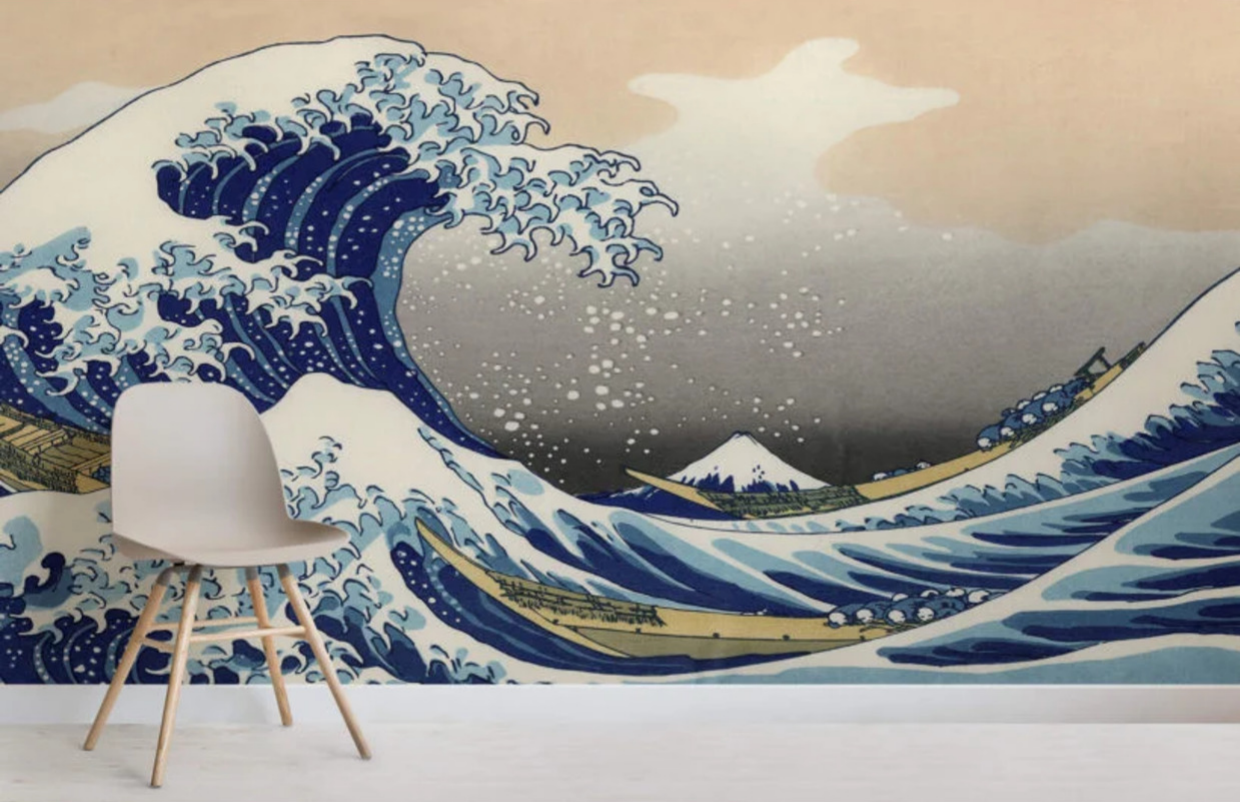}
\caption{As an illustration of a Humean mosaic, this figure shows the painting ``The Great Wave off Kanagawa'' by \cite{hokusai_1830} reproduced on a wall. On a canonical Humean view of laws (e.g., \cite{Lewis1973,Lewis1999}) one imagines that every bit of paint is completely independent of every other bit of paint (i.e., there are no necessary connections between distinct existences). By contrast, \citeauthor{Demarest2017}'s (\citeyear{Demarest2017}) view allows us to imagine that these metaphorical bits of paint have some causal powers over each other. On either view, the laws are to be read off of the mosaic via a Best Systems Analysis. Regarding Regularity Relationism (discussed in Sec.~\ref{SecDFGeometry}) one imagines peeling all of the paint off of the wall in a single topologically connected sheet. The fact that we can then freely bend and stretch this topological sheet shows us that the Humean mosaic does not require any geometric underpinnings.}\label{FigMosaic} 
\end{center}
\end{figure}

As a rejoinder, a proponent of the dynamic-second view of laws might complain that, sans metaphysically weighty laws, the Humeans now have no explanation whatsoever for dynamics. Isn't science supposed to (among other things) explain the dynamical behavior of matter? This is a valid complaint against the canonical Humean view since it does take dynamics to be unexplained (i.e., it accepts the orderliness of the Humean mosaic as a brute fact). This, however, is not a necessary feature of dynamics-first views. Indeed, another notable dynamics-first view of laws is the generalization of \citeauthor{Lewis1973}'s (\citeyear{Lewis1973,Lewis1999}) Best Systems Analysis (BSA) by \cite{Demarest2017}. This view allows for causal powers and dispositions to exist on the Humean mosaic, see Fig.~\ref{FigMosaic}. On such a view the laws are then the theorems/axioms of whichever theory best systematizes this arrangement of dispositional and categorical properties. Importantly, on \citeauthor{Demarest2017}'s (\citeyear{Demarest2017}) view, the orderliness of the Humean mosaic is not accepted as a brute fact; Instead, it is explained in terms of the causal powers inherent in matter itself. These causal powers are then accepted as primitive. As this example shows, dynamics-first views can explain dynamics in terms of yet-more fundamental dynamics (which itself remains unexplained).

It should be noted that \citeauthor{Demarest2017}'s (\citeyear{Demarest2017}) BSA abandons the canonical Humean idea that there are no necessary connections between distinct existences. As \cite{BhogalHarjit2020Halo} observes, however, her view is still Humean in the following ``unofficial'' sense:
\begin{quote}
\singlespacing\vspace{-0.5cm}
One common motivation [for Humeanism about laws] comes from the way that the defender of the BSA claims that their view mirrors scientific practice (e.g., \cite{Demarest2017}). The BSA sticks very closely to the actual practice of science—in fact, one way to describe the view is that it takes the actual methodology of science and it mirrors that methodology in the account of the metaphysics of laws. \cite{Hall2015} calls this the “unofficial guiding idea” behind Humeanism about laws.\footnote{\citeauthor{BhogalHarjit2020Halo} (\citeyear[pg. 8]{BhogalHarjit2020Halo})}
\end{quote}
\enlargethispage{1cm} That is, the common feature shared by \citeauthor{Demarest2017}'s (\citeyear{Demarest2017}) BSA and more canonical Humean views is exactly the above-discussed paralleling of the arrow of metaphysically explanation with the arrow of knowledge generation. As I will argue throughout this paper, the benefits of aligning these two arrows extend beyond the metaphysics of laws debate. All dynamics-first views make use of this alignment and are therefore ``unofficially'' Humean. In every context, this alignment is favorable since we are generally on a more secure footing when we explain the less-directly-known in terms of the more-directly-known rather than vice-versa. Isn't science supposed to (among other things) explain the world's structural features in terms of the dynamical behavior of matter?\footnote{Moreover, isn't science supposed to explain the non-fundamental dynamics of matter in terms of its more fundamental dynamics? Both here and in the ``structural'' question in the main text, by ``explanation'' I mean metaphysically supported explanation, see footnote~\ref{FnMetaExplain}.} 

Before moving on to the other three contexts allow me to make explicit how a capacity for nomological redescription can be leveraged in support of a dynamics-first view of laws. A distinction must first be made between axiom-like laws of nature and theorem-like laws (i.e., derived laws). Throughout this paper, my focus is exclusively on axiom-like laws since the dynamics-second views only seek to grant metaphysical weight to 
axiom-like laws. (Granting nomological force to theorem-like laws as well as axiom-like laws is clearly redundant). This focus on axiom-like laws should be contrasted with other parts of the metaphysics of laws literature. For instance, \cite{Lewis1973,Lewis1999} takes the laws to be (a subset of) the theorems of whichever theory best systematizes the Humean mosaic.

This brings us to the typical starting place for Humean views of laws: the Humean mosaic. Given a single Humean mosaic, one's task is typically to read the laws of nature off of it by creating a theory about this single mosaic from evidence available on this single mosaic. To my mind, however, proceeding in this way conflates two distinct processes: 1) the process of constructing an initial theory, and 2) the process of selecting the best axiom-like laws for this theory. For now, let us focus on the second process. Moreover, let us temporarily adopt a syntactic view of theories such that the axiom-like laws of a given axiomatized theory are simply (a subset of) the axioms of that theory. Now suppose that by some unspecified process one arrives at an empirically successful theory which is stated in first-order logic in terms of some (perhaps many, perhaps complicated) axioms. What capacity for re-axiomatization do we have such that we can find some better axioms for this theory?

First notice that re-axiomatizing a theory does not change any of its logical consequences. This merely amounts to a nomological redescription of our theory. Moreover, note that by applying standard deductive logic techniques one can access \textit{every possible} nomological description of one's theory. Given this overabundance of nomological descriptions, one finds some motivation to metaphysically deflate the laws and to focus instead on the theory's law-independent content (e.g., its models viewed as Humean mosaics). From here one can then search for a good axiomatization of the theory (i.e., a good nomological description). Viewed from this law-neutral perspective, whichever axiom-like laws one ultimately settles on will reflect nothing metaphysically weighty in the world beyond them being one particularly nice way (among other options) of codifying the dynamical behavior of matter.

It is important to note that past making this metaphysical claim about the laws, dynamics-first views of laws are allowed to vary regarding a wide variety of further epistemological questions about the laws. What role does the notion of laws of nature play in the initial process of theory construction? How do we first come to have a nomological picture of the world? Given our capacity for nomological redescription, what guides us in selecting the axiom-like laws for our theories? That is, what does the term ``particularly nice'' mean exactly? Is the law-selection process objective and mind-independent, making the laws real (despite them being metaphysically deflated)? Or, alternatively, is there some non-trivial element of conventionality to it? As I will discuss in Sec.~\ref{SecTheoryLevel}, certain answers to these epistemological questions may have interesting metaphysical consequences.

\subsection{Dynamics-First Views of Coordinates}\label{SecDFCoor}

Let us move on from the metaphysics of laws debate to the next context: coordinate systems. Suppose that after investigating the inertial behavior of a scalar field, $\varphi(t,x,y,z)$, and a point-like particle with trajectory, $X(\tau)=(t(\tau),x(\tau),y(\tau),z(\tau))$, one arrives at the following theory written in some fixed global coordinate system, $(t,x,y,z)$:
\begin{align}\label{CoordinateTheory1}
(\partial_t^2-\partial_x^2-\partial_y^2-\partial_z^2+M^2)
\, \varphi(t,x,y,z)&=0,\\
\label{CoordinateTheory2}
(\partial_\tau^2 t, \ \partial_\tau^2 x, \ \partial_\tau^2 y, \ \partial_\tau^2 z) &= F/m,\\
\label{CoordinateTheory3}
(\partial_\tau t)^2-(\partial_\tau x)^2-(\partial_\tau y)^2-(\partial_\tau z)^2&=1.
\end{align}
Eq.~\eqref{CoordinateTheory1} is the Klein-Gordon equation for a free scalar field with mass $M$. Eq.~\eqref{CoordinateTheory2} is the special-relativistic version of Newton's second law for a particle with a rest mass of $m$ undergoing a constant 4-force of $F=(f_t,f_x,f_y,f_z)$. In the $F=0$ case, it states that force-free particles travel on straight lines in the $(t,x,y,z)$-coordinate system. Eq.~\eqref{CoordinateTheory3} states that the particle is always moving slower than the speed of light.

A dynamics-second reading of the $(t,x,y,z)$-coordinate system is clearly inappropriate; This would involve granting some metaphysical weight to this coordinate system and then using it as an explanans for the inertial dynamics of $\varphi$ and $X$. This clearly will not work since coordinate systems are ultimately a representational artifact which we project onto the world. Indeed, it is nearly trivial to switch between a wide variety of different coordinate systems. Moreover, as I will soon demonstrate, we can always move into a coordinate-free formulation of our physics.\footnote{Or, at least, almost always, see~\cite{pittphilsci8895}.} From here we can then access \textit{every possible} coordinate description of our theory. These considerations push us strongly towards a dynamics-first view of coordinates: The coordinate systems which feature in our best theories reflect nothing metaphysically weighty in the world beyond them being a particularly nice way (among other options) of codifying the dynamical behavior of matter.

Hence, appealing to an inertial coordinate system will not give us a metaphysically supported explanation of the inertial behavior of matter. If anything is to explain the inertial behavior of matter, it would be the inertial structure of spacetime itself (this being the feature of the world which our inertial coordinate systems have latched onto). But where can this inertial structure be seen in the above theory? As I will now demonstrate, this theory's inertial structure will miraculously appear when we move to a generally covariant (i.e., a coordinate-independent) representation of our physics. (Readers unfamiliar with differential geometry need not worry, no technical know-how is needed to understand what follows.) Written in the coordinate-free language of differential geometry, Eq.~\eqref{CoordinateTheory1} -~\eqref{CoordinateTheory3} become,
\begin{align}
\label{NoCoorTheory1}
(\eta^\text{ab}\nabla_\text{a}\nabla_\text{b}-M^2)\varphi&=0,\\
\label{NoCoorTheory2}
\dot{X}^\text{a}\nabla_\text{a} \dot{X}^\text{b}&=F^\text{b}/m,\\
\label{NoCoorTheory3}
\eta_\text{ab}\dot{X}^\text{a}\dot{X}^\text{b}&=-1,
\end{align}
where $\varphi:\mathcal{M}\to\mathbb{R}$ is a real-valued scalar field on a spacetime manifold $\mathcal{M}\cong\mathbb{R}^4$. Here $\eta_\text{ab}$ is a fixed flat Lorentzian metric with signature $(-,+,+,+)$ defined on $\mathcal{M}$ and $\nabla_\text{c}$ is the unique covariant derivative compatible with $\eta_\text{ab}$, i.e., with $\nabla_\text{c}\eta_\text{ab}=0$. Here $X:\mathbb{R}\to\mathcal{M}$ is a particle trajectory with $\dot{X}^\text{a}:\mathbb{R}\to\text{T}\mathcal{M}$ being its tangent vector. Finally, $F^\text{b}$ is the 4-force which is being applied to the particle.

The key takeaway from rewriting our theory in this way is the appearance of the following three mathematical objects: a Lorentzian metric, $\eta_\text{ab}$, a covariant derivative, $\nabla_\text{c}$, and a smooth manifold, $\mathcal{M}$. Each of these three objects was in some sense ``hidden'' by our previous use of the $(t,x,y,z)$-coordinate system. We have now found this theory's inertial structure. In particular, Eq.~\eqref{NoCoorTheory2} still says that force-free particles travel in ``straight lines''. Now, however, rather than checking the straightness of the trajectory $X(\tau)$ against the $(t,x,y,z)$-coordinate system, we are instead evaluating its straightness with respect to a piece of fixed geometric structure, namely $\nabla_\text{c}$ (which happens to be uniquely determined by $\eta_\text{ab}$). This sets the standard of straightness on $\mathcal{M}$, and thereby picks out the theory's inertial structure.

\subsection{Dynamics-First Views of Geometry}\label{SecDFGeometry}

As I noted above, if anything is going to help us explain the inertial behavior of matter, it would be the inertial structure of the spacetime manifold itself (i.e., the Lorentzian metric, $\eta_\text{ab}$). This brings us to the next context for the dynamics-first vs dynamics-second debates: geometry. On a dynamics-second view of geometry, our above investigation of inertial dynamics has revealed to us the world's primitive geometric substructures. Moreover, these geometric substructures are then taken to be metaphysically weighty in such a way that we can use them to help explain the inertial behavior of matter. For instance, one might argue that the law ``force-free particles travel on straight lines'' presupposes the existence of some primitive geometric structures on $\mathcal{M}$ with respect to which the concept of ``straightness'' is defined.

In opposition to this dynamics-second explanatory strategy, we have the dynamics-first view of geometry championed by \cite{RBrown2005} and others.\footnote{The dynamics-first views of geometry are, roughly, the dynamical side of the geometrical vs dynamical spacetime debate. For a recent overview of the dynamics-first side of this debate see \cite{BrownRead2018} and references therein, especially \cite{RBrown2005,Nonentity,BrownPooley1999,Menon2019,HuggettNick2006TRAo,StevensSyman2014Tdat}.} According to this view, the dynamical behavior of matter does not require any geometric underpinnings. In fact, the arrow of explanation is here reversed: The Lorentzian metric, $\eta_\text{ab}$, is seen as being merely a codification of the dynamical behavior of measuring rods and clocks. As in the laws' context, making the arrows of knowledge and explanation parallel allows us to adopt a metaphysically deflated view of geometry and thereby avoid the following tricky metaphysical questions. How is a force-free particle supposed to ``know'' what a straight line is? Must it consult with some fixed geometrical background structure? How exactly does the geometry guide the matter's dynamics? Why can't the particles/fields push back on the geometry?\footnote{As a historical note, it was this last line of questioning which led Einstein to develop his theory of general relativity, see \cite{BrownHarveyR2013Etro}.} Such questions are raised against dynamics-second views throughout the dynamical vs geometrical spacetime debate, especially in the work of \cite{RBrown2005}.

This dynamics-first view of geometry bares more than just a superficial resemblance to the above-discussed dynamics-first view of laws. Indeed, one can pick out a theory's geometric structures in roughly the same way that Humeans pick out a theory's laws. As a particularly vivid example of this, consider the view of Regularity Relationism developed by \cite{HuggettNick2006TRAo} and later extended by \cite{Pooley2013} and \cite{StevensSyman2014Tdat}. The following imagery provides a sketch of this view. Imagine carefully peeling all of the metaphorical Humean paint off of the wall in one piece, see Fig.~\ref{FigMosaic}. Moreover, imagine that this detached paint sheet is locally bendable and stretchable but cannot tear or break apart. Now imagine searching for a particularly nice way of bending and stretching this detached paint sheet before reassociating it with the fixed geometry of the wall. (As in the laws' context, different dynamics-first views of geometry are allowed to vary regarding what qualifies as ``particularly nice''.)

The purpose of this metaphorical imagery is to show us that the Humean mosaic does not require any geometric underpinnings. Indeed, the fact that we can imagine freely bending and stretching this detached paint sheet shows that it has only a topological structure, i.e., no geometric structure. Moreover, our ability to play so freely with the relationship between the mosaic and the wall's fixed geometry suggests that we might be able to pick out the theory's geometric structure by something like a Best Systems Analysis. As I will argue in Sec.~\ref{SecRedescription}, the capacity for geometric redescription that Regularity Relationism gives us is analogous to our above-discussed capacity for nomological redescription and coordinate redescription. Namely, it gives us a geometry-neutral framing of our theory's models (i.e., the detached paint sheets) from which we can freely explore \textit{every possible} geometric description of our theory. These analogous redescription techniques push us towards the following dynamics-first view of geometry: The fixed geometric structures which feature in our best theories (e.g., $\eta_\text{ab}$ in special relativity) reflect nothing metaphysically weighty in the world beyond them being one particularly nice way (among other options) of codifying the dynamical behavior of matter.

Just as in the laws' context, past making this metaphysical claim, dynamics-first views of geometry are allowed to vary regarding a wide variety of further epistemological questions about geometry. Indeed, the questions we saw at the end of Sec.~\ref{SecDFLaws} apply here nearly unchanged. What role do geometric notions play in the initial process of theory construction? How do we first come to have a geometrical picture of the world? Given our capacity for geometric redescription, what guides us in selecting a geometric framing for our theories? That is, what does the term ``particularly nice'' mean exactly? Is the geometry-selection process objective and mind-independent, making the theory's geometry real (despite it being metaphysically deflated)? Or, alternatively, is there some non-trivial element of conventionality to it? Unfortunately, it is beyond the scope of this paper to discuss these questions any further.

\subsection{Dynamics-First Views of Topology}\label{SecDFTopology}
The above discussion has laid the groundwork for us to now make a novel extension of the dynamics-first vs dynamics-second debates into the context of spacetime topology. \cite{Norton2008} has already raised the challenge of making such an extension. He correctly notes that deflating a theory's geometric structures as we did in Sec.~\ref{SecDFGeometry}  still leaves its topological structure (i.e., the spacetime manifold itself, $\mathcal{M}$) intact as a serious candidate for ontological commitment. This has been called the ``problem of pre-geometry'':\footnote{For further discussion of the problem of pre-geometry, see~\cite{linnemann2021constructivists,Chen_2021,Menon2019}.} The dynamics-first approach to spacetime described above only applies to the geometric structure of spacetime and not to its topological structure. Indeed, the above discussion of Regularity Relationism seems to rely upon imagining the Humean mosaic as a topologically connected sheet. 

In light of Norton's complaint, it is natural to wonder whether the above-discussed dynamics-first views of laws, coordinates, and geometry can be extended to cover our theories' topological structures as well. Indeed as \citeauthor{Norton2008} (\citeyear[pg. 833]{Norton2008}) notes, ``One can certainly conceive more modest versions [of the dynamics-first approach to spacetime] that presume some spatiotemporal notions and not others, although the resulting constructivist view becomes less interesting the more spatiotemporal properties it assumes.'' (Note that Norton calls the followers of the dynamics-first approach constructivists.) Hence, if such a dynamics-first view of topology could be put forward, then proponents of dynamics-second perspectives (such as Norton) should rejoice in having a more interesting and unified position to argue against.

In just the same way, proponents of dynamics-first views (of laws, coordinates, and/or geometry) should rejoice about the prospects of a dynamics-first view of topology. While one could evaluate these four debates independently, there is much to be gained by putting them into close contact with each other and allowing one's view to equilibrate across these four contexts. In addition to the rather banal point that analogous situations ought to be treated analogously, adopting a unified metaphysical orientation (either dynamics-first or dynamics-second) allows one's positions in these debates to lend each other mutual support and share in their common motivations. These considerations drive us to search for a dynamics-first view of topology.

But how should we then proceed? A way forward has already been revealed by the above discussion. Indeed, as I have stressed above, the dynamics-first views of laws, coordinates, and geometry are each supported by our existing capacities for redescription. Hence, one way to achieve a dynamics-first view of topology would be to develop some analogous topological redescription techniques (ideally, something as powerful as our existing techniques for nomological, coordinate, and geometric redescription). To this end, I have recently developed such a suite of topological redescription techniques, namely the ISE Method, see \cite{IntroISE} and \cite{ISEEquiv}.

In particular, \cite{ISEEquiv} has evaluated the degree to which the topological redescription techniques introduced in \cite{IntroISE} are analogous to our existing capacities for nomological, coordinate, and geometric redescription. Sec.~\ref{SecISEOverview} will summarize the results of these two papers. For now, it suffices to state their conclusions. The ISE Method allows us to remove and replace the topological underpinnings of our spacetime theories just as easily as we can switch between different coordinate systems (or different axiom systems, or different geometric descriptions).

The purpose of this paper is to assess (and ultimately validate) the claim that these new topological redescription techniques can be used to support a dynamics-first view of topology which is analogous to the above-discussed dynamics-first views of laws, coordinates, and geometry. In direct analogy with these views, the central claim of the dynamics-first view of topology is as follows: The topological structure which features in our best theories (e.g., the spacetime manifold, $\mathcal{M}$) reflects nothing metaphysically weighty in the world beyond it being one particularly nice way (among other options) of codifying the dynamical behavior of matter. (As in the previous contexts, different dynamics-first views of topology will be allowed to vary regarding what qualifies as ``particularly nice''.)

\enlargethispage{1cm} Allow me to briefly spell out some of the philosophical consequences of this view. Recall from above that dynamics-first views deny the metaphysical weightiness of our theory's laws/geometry/topology by aligning the arrows of knowledge and explanation, see Fig.~\ref{FigDynamicsSecondFirst}c). Ultimately, this means that our theories' nomological/geometrical/topological structures are to be limited to the epistemological roles they play in helping us study and codify the dynamical behavior of matter. As I have discussed above, this allows us to avoid many tricky metaphysical questions. Whatever metaphysical questions about laws/geometry/topology remain are then reduced to epistemological questions about the roles these structures play in our theorizing. Importantly, however, taking a dynamics-first perspective does not obligate one to adopt any specific views about these further epistemological questions. Indeed, there are plenty of epistemological issues left on the table for dynamics-first views to disagree about.

Just as in the previous context, dynamics-first views of topology are allowed to give varying answers to a wide variety of epistemological questions about spacetime topology.\footnote{As \cite{STepistemology} notes, these epistemological questions about spacetime have recently been neglected in favor of more metaphysical questions about space and time.} Indeed, the questions we saw at the end of Sec.~\ref{SecDFLaws} and Sec.~\ref{SecDFGeometry} apply here nearly unchanged. What role do topological notions play in the initial process of theory construction? How do we first come to have a topological picture of the world? Given our capacity for topological redescription, what guides us in selecting a topological framing for our theories? That is, what does the term ``particularly nice'' mean exactly? Is the topology-selection process objective and mind-independent making the theory's topology real (despite it being metaphysically deflated)? Or, alternatively, is there some non-trivial element of conventionality to it? 

Despite being open on all of these epistemological points, the dynamics-first view of topology does present a definite metaphysical picture of space and time which is of a roughly Neo-Kantian character (contra Newton and Leibniz). Namely, according to the dynamics-first view of topology, rather than corresponding to any fundamental substances or relations out there in the world, the spacetime manifolds which feature in our best theories ought to be thought of as merely being an aspect of how we \textit{represent} the world. Compare this view with the following quote from Kant about the metaphysical nature of space:
\begin{quote}
\singlespacing\vspace{-0.5cm}
Space is not something objective and real, nor a substance, nor an accident, nor a relation; instead, it is subjective and ideal, and originates from the mind’s nature in accord with a stable law as a scheme, as it were, for coordinating everything sensed externally.\footnote{\citeauthor{KantDissertation} (\citeyear[pg. 397]{KantDissertation})} 
\end{quote}
That is, according to Kant space is not a substance (denying Newton) nor a relation (denying Leibniz) but rather something which our minds project onto the world as an organizing principle. Of course, Kant then goes on to make some further epistemological claims about space and time being matters of a priori intuition which are necessary for perception and cognition to even begin. The dynamics-first view of topology agrees with Kant's metaphysical claims about space and time but is not committed to his further epistemological claims. For instance, the dynamics-first view of topology might instead be taken in a conventionalist/Neo-Kantian direction paralleling~\citeauthor{Poincare}'s (\citeyear{Poincare}) conventionalism about geometry.\footnote{For some discussion of where Poincar\'e sits on the spectrum from Neo-Kantians to conventionalists, see~\cite{IVANOVA2015114}.} Alternatively, the dynamics-first view of topology could be taken in a realist direction paralleling realism about Humean laws of nature, see~\cite{Lewis1994}.

The remainder of this paper is organized as follows. Sec.~\ref{SecTheoryLevel} will highlight one key way in which dynamics-first views are allowed to vary: They might apply either at the model level or at the theory level. In Sec.~\ref{SecRedescription}, I will then stress how the dynamics-first views of laws, coordinates, and geometry are supported by our existing capacities for nomological, coordinate, and geometric redescription respectively. Sec.~\ref{SecISEFTop} will then present some analogous topological redescription techniques which can be leveraged in support of an analogous dynamics-first view of topology.

\section{Dynamics-First Views at the Theory Level and at the Model Level}\label{SecTheoryLevel}

\enlargethispage{-0.5cm} The previous section has highlighted that dynamics-first views are allowed to vary regarding a wide variety of epistemological questions. As I noted at the end of Sec.~\ref{SecDFLaws}, some of these epistemological questions may have metaphysical consequences. One such question deserves special attention. To what degree are our best theory's laws/geometry/topology either directly or indirectly associated with the actual world? Let us assume that the theory in question aims to represent the entirety of the actual world with one of its models (i.e., without the aid of its other models).\footnote{For some critiques of this assumptions, see~\cite{Curiel2020} and~\cite{pittphilsci19728}. Without this assumption the pressure to deliver a model-level understanding of our theory's laws/geometry/topology is greatly reduced.} Our question then becomes: To what degree are our theory's laws/geometry/topology identifiable within each of its models individually (i.e., without relying the relationships between models)?\footnote{For commentary on the importance of understanding the relationships between a theory's models see~\cite{Halvorson2016}.} Said differently, given any model of the theory is it possible for us to point to some structure within that model alone and declare, ``\textit{that} is what the theory's nomological, geometric, and/or topological structure has latched onto''? If so, let us say that we have a model-level understanding of the theory's laws, geometry, and/or topology. 

Importantly, dynamics-first views may or may not give us a model-level understanding of the theory's nomological, geometric, and/or topological structures. Indeed, as I will now discuss, the default position for dynamics-first views is that they do not. Recall that the dynamics-first vs dynamics-second debates are about whether or not certain theoretical structures (e.g., laws, geometry, topology, etc.) which appear in our best scientific theories should be granted any metaphysical weight in order to help us explain dynamics. What should be noticed about this framing of the debate is that these theoretical structures are just that, structures which appear in our theories. Hence, unless some further argument is provided, these structures should be thought of first and foremost as the laws/geometry/topology \textit{of the theory} and only indirectly the laws/geometry/topology \textit{of the world}. 

This indirect association of these structures with the actual world comes about simply because they appear in a theory in which is (or at least aspires to be) about the actual world. Namely, the theory in which these structures feature: 1) was built from real-world evidence, and 2) successfully makes predictions about real-world experiments. Moreover, if the theory is semantic then 3) its models may accurately represent parts of the actual world's matter and dynamics, or if the theory is syntactic then 4) its terms may successfully refer to parts of the actual world's matter and dynamics. These four points of connection with the actual world automatically give us a theory-level understanding of the theory's laws, geometry, and topology. Past this, additional argumentation must be provided to achieve a model-level understanding of these structures. As I will now discuss, it is possible (under certain assumptions) to achieve a model-level understanding of at least the nomological structures of a certain kind of theories.

To demonstrate this possibility, let us adopt a syntactic view of theories such that the axiom-like laws of theory are simply the axioms of that theory. Note that the theory's laws are axioms and therefore are sentences. If we want the truth-makers of these sentences at a given model to be contained strictly within that model, then we need to assume that the theory's axioms are non-modal (e.g., `all F's are G's'). Notably, this is not the case on~\citeauthor{Demarest2017}'s (\citeyear{Demarest2017}) account since her laws are allowed to refer to dispositions (e.g., `this would have happened if that had happened'). Under our non-modal assumption, however, each axiom-like law reports a pattern (e.g., the fact that all F's are G's) which is realized within each of the theory's models individually (i.e., independent of what is happening in its other models). If the theory's axioms are true at the actual world, then we can point to the corresponding pattern and declare, ``\textit{that} is what the theory's nomological structure has latched onto''. This gives us a model-level understanding of the laws of nature, as desired.

It should be clear from the above discussion how relaxing the modal assumption would seriously challenge our ability to have a model-level understanding of the laws of nature.\footnote{One might, however, follow \cite{Demarest2017} in claiming that the theory's nomological structure has latched onto patterns in the actual world's dispositional properties. Accepting such modal properties into one's ontology challenges our above assumption that our theories aim to represent the actual world with \textit{exactly one} of its models in isolation.} Moreover, as I will now discuss, such challenges also arise if we move to a semantic understanding of theories. On a semantic view of theories, the laws of nature are not pattern-reporting sentences. Instead, they are axiom-like rules for generating models, e.g., the laws are the equations to which the theory's models are solutions. Unlike in the syntactic case, it is possible that a semantic theory's laws of nature cannot be understood individually or assigned truth-values individually, see~\citeauthor{Ott2022} (\citeyear[Ch. 11.4]{Ott2022}). For instance, it is notoriously difficult to understand Newton's three laws of motion in such an individuated way, see~\citeauthor{Ott2022} (\citeyear[Ch. 3.2]{Ott2022}) and~\citeauthor{read_2023} (\citeyear[Ch. 1]{read_2023}). Indeed, a theory's laws might work together in non-trivial ways to generate the theory's models. Given that such a web of laws is not just a collection of pattern-reporting sentences, it is much more difficult to point to some concrete structure in either its models or in the actual world which this web of laws has latched onto. Fortunately, despite the difficulties we face in understanding such a web of laws at the model level, we can always understand them at the theory level (and thereby maintaining their indirect connection with the actual world).

The same theory-level vs model-level distinction can be applied in the context of geometry. Let us see how. Recall from Sec.~\ref{Sec2v1} the imagery associated with Regularity Relationism. Recall that one first imagines peeling all of the metaphorical Humean paint off of the Humean mosaic as a single topologically connected sheet, see Fig.~\ref{FigMosaic}. One then searches for a particularly nice way of bending and stretching this detached paint sheet before reassociating it with the fixed geometry of the wall. The way that I have described this process makes it sounds as if the theory's geometric structure is to be read off of a single Humean mosaic (i.e., a single model of our theory). Indeed, this is the typical way of understanding Regularity Relationism and would amount to a model-level understanding of the theory's geometric structure. As in the laws' context, however, this is not a necessary feature of dynamics-first views. 

Indeed, one's selection criteria for picking out the theory's geometric structure might require that it nicely codifies not just one of the theory's models/mosaics but multiple of its models/mosaics simultaneously. That is, one could search for a particularly nice way of simultaneously bending and stretching all of the detached paint sheets collectively. In particular, motivated by \cite{EarmanJohn1989Weas}, one could search for a geometric codification of the theory which captures the symmetry relationships which hold between the theory's models. Note that a theory's symmetries are not visible within its individual models; Instead, they are about certain relationships holding between the theory's models. In sum, one can easily imagine applying Regularity Relationism at a theory level. Even if it is only understood at a theory level, there is still a significant (albeit indirect) connection between the theory's geometric structure and the actual world. Namely, these geometric structures still feature in a theory which is about the actual world. This theory: 1) was built from real-world evidence, and 2) successfully makes predictions about real-world experiments, and 3) its geometry-neutral models may, in fact, accurately represent the actual world's matter and dynamics.

Some commentary on how the theory-level vs model-level distinction applies in the context of topology will be given at the end of Sec.~\ref{SecISEFTop}. Indeed, there I will put forward a dynamics-first view of topology which can likely only offer us a theory-level understanding of the theory's spacetime manifold.

\section{How Strong Redescriptive Capacities Support\\ Dynamics-First Views}\label{SecRedescription}

In Sec.~\ref{Sec2v1} I stressed that the dynamics-first views of laws, coordinates, and geometry are each supported by the existence of certain strong redescription techniques. In this section, I will elaborate on how analogous these various redescription techniques are and how they provide analogous support for their respective dynamics-first views. I do this in anticipation of putting forward the dynamics-first view of topology in Sec.~\ref{SecISEFTop}. Indeed, it will be also supported by some strong topological redescription techniques, namely the ISE Method.

A first point of analogy is that all three redescription techniques can be implemented in a broadly subject-neutral way. Namely, the mechanical process of re-axiomatization can be applied to our theories independently of what subject matter they are about: indeed, logic is subject-neutral (or at least it aspires to be). This is ultimately the reason why both \cite{Lewis1973,Lewis1999} and \cite{Demarest2017} can invoke Best Systems Analyses despite disagreeing radically on the metaphysical details of the to-be-codified mosaic; The logical process of re-axiomatization is metaphysics-blind. Similarly, the mechanical process of changing coordinates applies to our theories in exactly the same way no matter what the theory is about: e.g., scalar fields, particle trajectories, bosons, fermions, strings, etc. Finally, note that the paint-peeling imagery associated with Regularity Relationism can be applied equally well independent of the metaphysical thickness/thinness of the Humean paint (e.g., whether it contains dispositions or not). 

Indeed, unless one carefully directs one's effort to redescribe one's theory in a perspicuous way, then the resulting description is likely to be tremendously complicated. Almost all coordinate systems are horribly complicated; Almost all axiomatizations are horribly complicated, resulting in laws which are ill-suited to capture the theory's metaphysics; Almost all ways of bending and stretching the Humean mosaic result in geometric structures which are ill-suited to the theory in question. Said differently, each of these redescription techniques vastly overgenerates candidate descriptions (most of which will be tremendously complicated).

At this point, one may begin to worry about the overabundance of nomological, coordinate, and geometric descriptions which are on offer. In each case, however, our saving grace is that our redescriptive capacities are not only of wide scope but also complete in the following sense: By using these redescription techniques we can access \textit{every possible} nomological/coordinate/geometric description of our theory. Hence, even though descriptions will be horribly complicated, if there exists a particularly nice description, then we can find it. Note that this claim holds no matter what selection criteria one adopts regarding what qualifies as a ``particularly nice'' nomological/coordinate/geometric description.

Considering our strong capacities for nomological, coordinate, and geometric redescription, one might grow suspicious of any metaphysical weight being granted to these axioms systems, coordinate systems, or fixed geometric structures. That is, one might be led towards a dynamics-first view of laws, coordinates, and/or geometry. Indeed, given these strong capacities for redescription, one finds some motivation to focus on our theory's coordinate-independent, law-independent, and/or geometry-independent content. As I have discussed already in Sec.~\ref{SecDFCoor}, we can always move into a coordinate-neutral framing of our theory from which we can search for perspicuous coordinate systems. Similarly, our nomological redescription techniques allow us to conceive of our theories in a law-independent way: namely, we can view them as either a collection of sentences or as a collection of models. Moreover, each model of our theory can be thought of in an law-independent way by thinking of it as a possible Humean mosaic, see Fig.~\ref{FigMosaic}. Notice also that we can think of these Humean mosaics in a geometry-independent way by considering the detached paint sheets which result from the above-described paint-peeling process.

In sum, our nomological, coordinate, and geometric redescription techniques give us access to a law-neutral/coordinate-neutral/geometry-neutral description of our theories from which we can search for good nomological/coordinate/geometric descriptions. In each case, these strong redescription techniques lend support to the dynamics-first views of laws, coordinates, and geometry which were discussed in Sec.~\ref{Sec2v1}: The laws, coordinates, and/or fixed geometric structures which feature in our best theories reflect nothing metaphysically weighty in the world beyond them being one particularly nice way (among other options) of codifying the dynamical behavior of matter.

\section{My Dynamics-First View of Topology}\label{SecISEFTop}

The previous two sections have laid the groundwork for the dynamics-first view of topology which I will put forward in this section. As I have stressed above, the dynamics-first views of laws, coordinates, and geometry are supported by our capacities for nomological, coordinate, and geometric redescription respectively. My dynamics-first view of topology will be supported by some analogous topological redescription techniques. Namely, it will be supported by the ISE Method which I have developed in~\cite{IntroISE} and~\cite{ISEEquiv}.

\subsection{A Quick Overview of the ISE Method}\label{SecISEOverview}

This section will provide an overview of the ISE Method's topological redescription techniques. For details beyond what follows, I direct the interested reader to ~\cite{IntroISE} and~\cite{ISEEquiv}. The ISE Method is a three-step process: Internalize, Search, Externalize (hence the initials, ISE). In the internalization step, one first divorces the dynamical behavior of matter from its topological underpinnings. This divorce is carried out in a similar spirit to spacetime algebraicism, albeit with some key differences.\footnote{Roughly put, spacetime algebraicism is the view that
the dynamical behavior of matter (specifically, field states) does not require any topological underpinnings (specifically, because they can be thought of algebraically). By contrast, my view drops these parentheticals. For further discussion of spacetime algebraicism, see
\cite{Chen_2021,Menon2019,linnemann2021constructivists,Rynasiewicz,EarmanJohn1989Weas,ROSENSTOCK2015309}. Notably, \cite{Geroch1972} uses Einstein Algebras to reconstruct general relativity in this way.} For instance, the internalization process is similar to the move commonly made in non-relativistic quantum mechanics from a position representation of the wavefunction, $\psi(t,x)$, to a Hilbert space representation, $\ket{\psi(t)}$. One key difference, however, is that we here internalize time as well as space. Moreover, the ISE Method does not require the theory in question to have any algebraic structure. For instance, the ISE Method is applicable to theories about particle trajectories which lack any meaningful algebraic operations (e.g., no sums and products).

\enlargethispage{-0.5cm} Internalization results in a topology-neutral characterization (i.e., a spacetime-neutral characterization) of the theory in question. One next searches for the building blocks of a new spacetime framing of the theory. I call these building blocks pre-spacetime translation operations, or PSTOs. A technical definition of PSTOs can be found in~\cite{IntroISE}, but the following characterization is sufficient for our purposes. Let $S_\text{neutral}^\text{kin}$ be the theory's set of kinematically allowed models understood in a topology-neutral way. Picking a set of PSTOs involves picking out a pair of Lie groups, $G_\text{trans}$ and $G_\text{fix}\subset G_\text{trans}$, which act on $S_\text{neutral}^\text{kin}$. To qualify as PSTOs, this pair of Lie group actions on $S_\text{neutral}^\text{kin}$ need to be compatible with each other in two ways: they must have faithful construction compatibility and injection compatibility, for details see~\cite{IntroISE} and~\cite{ISEEquiv}. These compatibility constraints are sufficient to guarantee that our PSTOs are, in a certain sense, structurally indistinguishable from spacetime translations. An intuitive example would be translations, not in spacetime, but rather in old theory's Fourier space. 

Finally, in the externalization step, one uses whichever PSTOs one has chosen to construct a new topological setting for the theory in question. One does this simply by taking these PSTOs seriously as spacetime translations. Indeed, by construction, one's chosen pre-spacetime translation operations will end up becoming honest-to-goodness spacetime translations in the new theory. More concretely, the process goes roughly as follows. One first declares that the Lie group, $G_\text{trans}$, is to be a group of diffeomorphisms which act transitively on new spacetime, $\mathcal{M}_\text{new}$. Secondly, one declares that $G_\text{fix}$ is to be the stabilizer subgroup of $G_\text{trans}$'s action on $\mathcal{M}_\text{new}$. Making these two declarations uniquely specifies $\mathcal{M}_\text{new}$ up to diffeomorphism as $\mathcal{M}_\text{new}\cong G_\text{trans}/G_\text{fix}$. For details see Appendix A of~\cite{IntroISE}. In the above-discussed Fourier example, the new theory's spacetime is effectively a copy of the old theory's Fourier space.

To complete the externalization process, one then must map the old theory's states and dynamics onto $\mathcal{M}_\text{new}$. As~\cite{IntroISE} discusses, this final stage of the externalization process is tightly constrained by our choice of PSTOs. For instance, in the Fourier example, the new theory's states and dynamics are required to be the Fourier transform of the old theory's states and dynamics (up to some rescaling freedom). \cite{IntroISE} discusses this Fourier redescription example at length in the context of non-relativistic quantum mechanics where it amounts to a position-momentum duality.

The exact same topological redescription techniques were also used in~\cite{IntroISE} to redescribe a lattice theory (i.e., a theory set on a discrete spacetime, $\mathcal{M}_\text{old}\cong\mathbb{R}\times\mathbb{Z}$) as existing on a continuous spacetime manifold, $\mathcal{M}_\text{new}\cong\mathbb{R}\times\mathbb{R}$. This topological redescription is facilitated by the fact that the discrete spacetime theory in question admits a set of continuous pre-spacetime translation operations (see Fig. 1 in~\cite{IntroISE}). Moreover, ~\cite{ISEEquiv} has used the ISE Method to implement a Kaluza-Klein dimensional reduction (and to then undo this dimensional reduction). 

\cite{ISEEquiv} also discusses how the ISE Method can be applied outside of the scalar field case. Indeed, it can also be applied to theories about a parameterized particle trajectory, \mbox{$x:\mathcal{V}\to\mathcal{M}$}. For instance, consider a theory about a point-like particle, \mbox{$x_\text{old}:\mathcal{V}_\text{old}\to \mathcal{M}_\text{old}$}, moving on a M\"{o}bius strip over time (with $\mathcal{V}_\text{old}\cong\mathbb{R}$ and $\mathcal{M}_\text{old}\cong\mathbb{R}\times\mathbb{M}$). One can use the ISE Method to redescribe such a theory as being about an extended particle, \mbox{$x_\text{new}:\mathcal{V}_\text{new}\to \mathcal{M}_\text{new}$}, moving on the Euclidean plane over time (with $\mathcal{V}_\text{new}\cong\mathbb{R}^2$ and $\mathcal{M}_\text{new}\cong\mathbb{R}\times\mathbb{R}^2$).

Moving beyond these example applications,~\cite{ISEEquiv} has proved the ISE Equivalence Theorem which states the following. The ISE Method gives us access to effectively \textit{every possible} spacetime framing of a given theory's kinematical and dynamical content, bounded only by a weak spacetime-kinematic compatibility condition. This proves that the ISE Method is a completely general tool for topological redescription (at least, within the realm of spacetime theories which satisfy this spacetime-kinematic compatibility condition). Every instance of topological redescription between any two such theories can be implemented by the ISE Method. As \citeauthor{ISEEquiv} (\citeyear[Sec. 2.5]{ISEEquiv}) has discussed, this spacetime-kinematic compatibility is an extremely weak assumption. Hence, the scope of topological redescriptions offered to us by the ISE Method is extremely broad. 

\subsection{The ISE Method Supports a Dynamics-First View of Topology}\label{SecISESupport}

The ISE Method supports a dynamics-first view of topology in the same way that our existing capacities for nomological, coordinate, and geometric redescription respectively support dynamics-first views of laws, coordinates, and geometry. Indeed, in \cite{ISEEquiv} I already explored in depth the analogy between our existing capacities for redescription and the capacity for topological redescription offered to us by the ISE Method. I will now briefly review the conclusions of \cite{ISEEquiv} regarding the strength of this analogy. 

Like each of the above-discussed redescription techniques, the ISE Method can be applied to a wide range of theories in a broadly subject-neutral way. Namely, it does not matter whether the theory is about scalar fields or particle trajectories. Nor does it matter whether the theory's dynamics are linear or non-linear. Another mark of similarity is that if applied in an unguided way, the ISE Method radically overgenerates possible topological redescriptions of the theory in question. See the above-discussed ISE Equivalence Theorem. Indeed, just like our other redescription techniques almost all of the topological redescriptions on offer will be horribly complicated and ill-suited to the theory in question. Fortunately, however, these topological redescription techniques are also complete in the following sense. Under the assumption that any ``particularly nice'' topological description of a theory will have spacetime-kinematic compatibility, we have the following result. We are guaranteed that if there exists a particularly nice topological framing of one's theory, then it is accessible to us via the ISE Method.

The above-discussed points have already given us a strong analogy between the ISE Method and our existing capacities for coordinate, nomological, and geometric redescription. The similarities do not stop here. Indeed, as I noted above, the ISE Method offers us topological redescriptions of the theory in question by first moving into a topology-neutral characterization of this theory. In light of our strong capacity for topological redescription, one might grow suspicious of any metaphysical weight being granted to the topological structures which appear in our theories (e.g., the spacetime manifold). That is, one is led towards a dynamics-first view of topology: The topological structure which features in our best theories (e.g., the spacetime manifold) reflects nothing metaphysically weighty in the world beyond it being one particularly nice way (among other options) of codifying the dynamical behavior of matter.

In closing, one final point should be noted regarding the ISE Method. The way in which the ISE Method generates new topological settings for our theories likely forces us to have a theory-level understanding of our theory's topological structure (i.e., its spacetime manifold). To see this, recall from Sec.~\ref{SecISEOverview} that the theory's new spacetime manifold, $\mathcal{M}_\text{new}$, is built out of a pair of groups which act on $S_\text{neutral}^\text{kin}$, i.e., the theory's set of kinematically allowed models. For any given model, $\sigma\in S_\text{neutral}^\text{kin}$, the subset of models which are involved in this construction contains at least the orbit of the Lie group $G_\text{trans}$ at $\sigma$. This reliance upon multiple models in the construction of a new spacetime setting suggests that it will be difficult to achieve a model-level understanding of the theory's topological structure. 

That is, given any spacetime-neutral model of the theory, it is unlikely that we will be able to point to some structure within this model alone and declare, ``\textit{that} is what the theory's topological structure has latched onto''. As I have discussed in Sec.~\ref{SecTheoryLevel}, however, there is still a significant (albeit indirect) connection between the theory's topological structure and the actual world. Namely, the theory's topological structure still features in a theory which is about the actual world. In particular, this theory: 1) was built from real-world evidence, and 2) successfully makes predictions about real-world experiments, and 3) its spacetime-neutral models may, in fact, accurately represent the actual world's matter and dynamics.

\section{Conclusion}\label{SecConclusion}

This paper has unified certain aspects of the metaphysics of laws debate and the geometrical vs dynamical spacetime debate by framing them as dynamics-first vs dynamics-second debates. I have then extended these debates outside of their original contexts of laws and geometry into the context of spacetime topology. As I discussed in Sec.~\ref{Sec2v1}, in each context, the central issue in each of these debates is whether or not one should grant any metaphysical weight to various theoretical structures (e.g., laws, geometry, topology, etc.) in hopes of giving a metaphysically supported explanation of dynamics.

I take it as a given that the theories in which these nomological, geometric, and topological structures appear are produced by studying and codifying the dynamical behavior of matter. That is, I take both the dynamics-first and the dynamics-second camps to agree upon the direction of the arrow of knowledge, see Fig.~\ref{FigDynamicsSecondFirst}. After this point of epistemological agreement, however, these two camps have a metaphysical disagreement about the direction of the arrow of explanation. In particular, the dynamics-second views take the presence of certain nomological, geometric, and/or topological structures in our best theories to reveal to us aspects of the world's metaphysical substructure. Moreover, they grant these nomological, geometric, and/or topological substructures enough metaphysical weight such that we can then turn around and use them to help explain the very dynamics which we began by investigating, see Fig.~\ref{FigDynamicsSecondFirst}b).

As I have discussed in Sec.~\ref{Sec2v1}, some tricky metaphysical questions arise for dynamics-second views. One way to avoid these questions (and to achieve a sparser metaphysics) is to reverse the arrow of explanation, putting dynamics first so to speak. As I have discussed in Sec.~\ref{Sec2v1}, adopting this explanatory strategy gives dynamics-first views an ``unofficially'' Humean character even outside of the laws' context. In particular, dynamics-first views deny the metaphysical weightiness of our theory's laws, geometry, and/or topology by aligning the arrow of explanation with the arrow of knowledge, see Fig.~\ref{FigDynamicsSecondFirst}c). Ultimately, this means that these structures are to be limited to the epistemological roles they play in helping us to study and codify the dynamical behavior of matter. This maneuver avoids many tricky metaphysical questions about these structures and subsequently reduces whatever metaphysical questions remain to epistemological questions regarding the roles these structures play in our theorizing. Importantly, however, taking a dynamics-first perspective does not obligate one to adopt any specific views about these further epistemological questions. Indeed, as I have discussed in Sec.~\ref{SecDFLaws}, Sec.~\ref{SecDFGeometry}, and Sec.~\ref{SecDFTopology}, there are plenty of epistemological issues left on the table for dynamics-first views to disagree about (some of which may have metaphysical consequences, see Sec.~\ref{SecTheoryLevel}). 

The main goal of this paper, however, was to introduce a dynamics-first view of topology in Sec.~\ref{SecISEFTop}. In Sec.~\ref{SecRedescription}, I stressed how the dynamics-first views of laws, coordinates, and geometry are respectively supported by our existing capacities for nomological, coordinate, and geometric redescription. Sec.~\ref{SecISEOverview} then introduced some powerful new topological redescription which I developed in~\cite{IntroISE} and~\cite{ISEEquiv}, namely the ISE Method. As I discussed in Sec.~\ref{SecISESupport}, these new topological redescription techniques are analogous to our existing capacities for nomological, coordinate, and geometric redescription. Hence, they give analogous support to an analogous dynamics-first view of topology: The topological structure which features in our best theories (e.g., the spacetime manifold) reflects nothing metaphysically weighty in the world beyond it being one particularly nice way (among other options) of codifying the dynamical behavior of matter. As I discussed at the end of Sec.~\ref{SecDFTopology}, the metaphysics of space and time put forward by this view has a notably Neo-Kantian character (contra Newton and Leibniz) while remaining open regarding Kant's further epistemological claims. Indeed, the dynamics-first view of topology could alternatively be taken in either a conventionalist/Neo-Kantian direction (paralleling~\cite{Poincare}) or a realist direction (paralleling~\cite{Lewis1994}).

\singlespacing
\bibliographystyle{dcu}
\bibliography{references}

@incollection{Friedman1983,
    author = {Michael Friedman},
    editor = {ibid.},
    booktitle = {Foundations of Space-Time Theories: Relativistic Physics and Philosophy of Science},
    publisher = {Princeton University Press},
    title = {Space-Time Theories},
    year = {1983},
    chapter = {2},
    pages = {32-70},
}

@book{EarmanJohn1989Weas,
publisher = {MIT Press},
isbn = {9780262050401},
year = {1989},
title = {World enough and space-time: {A}bsolute versus relational theories of space and time},
author = {Earman, John},
lccn = {LC},
}

@article{Norton2008,
issn = {0007-0882},
journal = {The British Journal for the Philosophy of Science},
pages = {821--834},
volume = {59},
publisher = {Oxford University Press},
number = {4},
year = {2008},
title = {Why Constructive Relativity Fails},
author = {Norton, John D.},
}

@article{Menon2019,
issn = {0031-8248},
journal = {Philosophy of science},
pages = {1273--1283},
volume = {86},
publisher = {The University of Chicago Press},
number = {5},
year = {2019},
title = {Algebraic Fields and the Dynamical Approach to Physical Geometry},
copyright = {Copyright 2019 by the Philosophy of Science Association. All rights reserved.},
language = {eng},
address = {Cambridge},
author = {Menon, Tushar},
}

@Article{Geroch1972,
author={Geroch, Robert},
title={Einstein algebras},
journal={Communications in Mathematical Physics},
year={1972},
month={12},
day={01},
volume={26},
number={4},
pages={271-275},
issn={1432-0916},
doi={10.1007/BF01645521},
}

@article{Nonentity,
issn = {1871-1774},
journal = {Philosophy and Foundations of Physics},
volume = {1},
year = {2006},
title = {Minkowski Space-Time: A Glorious Non-Entity},
language = {eng},
author = {Brown, Harvey R. and Pooley, Oliver},
pages = {67-89},
}

@incollection{BrownPooley1999,
    author = {Brown, Harvey R. and Pooley, Oliver},
    title = {The origins of the spacetime metric: Bell’s {L}orentzian pedagogy and its significance in general relativity},
    publisher = {Cambridge University Press},
    booktitle = {Physics Meets Philosophy at the Planck Scale},
    editor = {Callender, Craig and Huggett, Nick},
    isbn = {052166280X},
    year = {2001},
    editors = {Callender, Craig and Huggett, Nick},
    chapter = {15},
    pages = {256--72},
}

@phdthesis{StevensSyman2014Tdat,
author = {Stevens, Syman},
title = {The dynamical approach to relativity as a form of regularity relationalism},
    school = {University of Oxford},
    year = {2014}
}

@article{HuggettNick2006TRAo,
issn = {0026-4423},
journal = {Mind},
pages = {41--73},
volume = {115},
publisher = {Oxford University Press},
number = {457},
year = {2006},
title = {The Regularity Account of Relational Spacetime},
author = {Huggett, Nick},
}

@incollection{Pooley2013,
publisher = {Oxford University Press},
booktitle = {The Oxford Handbook of Philosophy of Physics},
editor = {Batterman, Robert},
chapter = {15},
isbn = {9780195392043},
year = {2013},
title = {Substantivalist and Relationalist Approaches to Spacetime},
pages = {522-586},
language = {eng},
author = {Pooley, Oliver},
}

@article{DoratoMauro2007RTbS,
issn = {0269-8595},
journal = {International studies in the philosophy of science},
pages = {95--102},
volume = {21},
publisher = {Routledge},
number = {1},
year = {2007},
title = {Relativity Theory between Structural and Dynamical Explanations},
language = {eng},
author = {Dorato, Mauro},
}

@incollection{Demarest2017,
publisher = {Oxford University Press},
booktitle = {Causal Powers},
isbn = {9780198796572},
year = {2017},
title = {Powerful Properties, Powerless Laws},
author = {Demarest, Heather},
editor = {Jonathan D. Jacobs},
chapter = {4},
pages = {38-54},
}

@book{Lewis1973,
	author = {David Lewis},
	title = {Counterfactuals},
	year = {1973},
	publisher = {Blackwell Publishers}
}

@article{Lewis1994,
	volume = {103},
	journal = {Mind},
	number = {412},
	publisher = {Oxford University Press},
	title = {Humean Supervenience Debugged},
	year = {1994},
	author = {David Lewis},
	doi = {10.1093/mind/103.412.473},
	pages = {473--490}
}

@book{Lewis1999,
booktitle = {Papers in Metaphysics and Epistemology},
isbn = {9780511625343},
year = {1999},
title = {Papers in Metaphysics and Epistemology},
language = {eng},
publisher = {Cambridge University Press},
author = {Lewis, David},
}

@article{BhogalHarjit2020Halo,
author = {Bhogal, Harjit},
title = {Humeanism about laws of nature},
journal = {Philosophy Compass},
volume = {15},
number = {8},
pages = {e12696},
issn = {1747-9991},
doi = {https://doi.org/10.1111/phc3.12696},
eprint = {https://compass.onlinelibrary.wiley.com/doi/pdf/10.1111/phc3.12696},
year = {2020}
}

@InCollection{sep-laws-of-nature,
	author       =	{Carroll, John W.},
	title        =	{{Laws of Nature}},
	booktitle    =	{The {Stanford} Encyclopedia of Philosophy},
	editor       =	{Edward N. Zalta},
	howpublished =	{\url{https://plato.stanford.edu/archives/win2020/entries/laws-of-nature/}},
	year         =	{2020},
	publisher    =	{Metaphysics Research Lab, Stanford University}
}

@book{BirdTextbook,
    author = {Bird, Alexander},
    title = "{Nature's Metaphysics: Laws and Properties}",
    publisher = {Oxford University Press},
    year = {2007},
    month = {06},
    isbn = {9780199227013},
    doi = {10.1093/acprof:oso/9780199227013.001.0001},
}

@article{LewisDavid1983Nwfa,
issn = {0004-8402},
journal = {Australasian journal of philosophy},
pages = {343--377},
volume = {61},
publisher = {Taylor & Francis Group},
number = {4},
year = {1983},
title = {New work for a theory of universals},
language = {eng},
author = {Lewis, David},
}

@article{BusinessOfLaws,
issn = {0012-2017},
journal = {Dialectica},
pages = {577--588},
volume = {70},
number = {4},
year = {2016},
title = {It is the Business of Laws to Govern},
author = {Schaffer, Jonathan},
}

@book{RBrown2005,
	year = {2005},
	publisher = {Oxford University Press},
	title = {Physical Relativity: Space-Time Structure From a Dynamical Perspective},
	author = {Harvey R. Brown}
}

@book{MaudlinTim2012Pop:,
publisher = {Princeton University Press},
booktitle = {Philosophy of physics: {S}pace and time},
isbn = {9781400842339},
year = {2012},
title = {Philosophy of physics: {S}pace and time},
author = {Maudlin, Tim},
}

@incollection{BrownRead2018,
  author = {Harvey R. Brown and James Read},
  title     = {The dynamical approach to spacetime theories},
  booktitle = {The Routledge companion to philosophy of physics},
  publisher = "Routledge",
  year      = {2022},
  editor    = {Knox, Eleanor and Wilson, Alastair},
  chapter   = {6},
  pages = {70-85},
}

@incollection{Hall2015,
author = {Hall, Ned},
publisher = {John Wiley and Sons, Ltd},
isbn = {9781118398593},
title = {Humean Reductionism about Laws of Nature},
booktitle = {A Companion to David Lewis},
chapter = {17},
pages = {262-277},
editor = {Loewer, Barry and Schaffer, Jonathan and Beebee, Helen},
doi = {https://doi.org/10.1002/9781118398593.ch17},
eprint = {https://onlinelibrary.wiley.com/doi/pdf/10.1002/9781118398593.ch17},
year = {2015},
}

@misc{hokusai_1830, title={Under the Wave off {K}anagawa [Painting]}, author={Hokusai, Katsushika}, year={1830}}

@article{Chen_2021,
	doi = {10.1016/j.shpsa.2021.08.011},
	year = 2021,
	month = {8},
	publisher = {Elsevier {BV}},
	volume = {89},
	pages = {188--201},
	author = {Lu Chen and Tobias Fritz},
	title = {An algebraic approach to physical fields},
	journal = {Studies in History and Philosophy of Science Part A}
}

@article{ROSENSTOCK2015309,
title = {On {E}instein algebras and relativistic spacetimes},
journal = {Studies in History and Philosophy of Science Part B: Studies in History and Philosophy of Modern Physics},
volume = {52},
pages = {309-316},
year = {2015},
issn = {1355-2198},
doi = {https://doi.org/10.1016/j.shpsb.2015.09.003},
author = {Sarita Rosenstock and Thomas William Barrett and James Owen Weatherall}
}

@incollection{BrownHarveyR2013Etro,
    author = {Brown, Harvey R. and Lehmkuhl, Dennis},
    title = {Einstein, the reality of space, and the action-reaction principle},
    booktitle = {Einstein, Tagore, and the nature of reality},
    editor = {Ghose, P},
    publisher = {Routledge},
    year = {2020}, 
    chapter = {1},
    pages = {1-20},
}

@book{Ott2022,
year = {2022},
title = {The metaphysics of laws of nature: {T}he rules of the game},
language = {eng},
publisher = {Oxford University Press},
author = {Ott, Walter},
}

@book{ArmstrongNomNec,
place={Cambridge},
series={Cambridge Studies in Philosophy},
title={What is a Law of Nature?},
DOI={10.1017/CBO9781139171700},
publisher={Cambridge University Press},
author={Armstrong, David M.},
year={1983},
}

@article{JANSSEN200926,
title = {Drawing the line between kinematics and dynamics in special relativity},
journal = {Studies in History and Philosophy of Science Part B: Studies in History and Philosophy of Modern Physics},
volume = {40},
number = {1},
pages = {26-52},
year = {2009},
issn = {1355-2198},
doi = {https://doi.org/10.1016/j.shpsb.2008.06.004},
author = {Michel Janssen}
}

@incollection{Ott2009,
    author = {Ott, Walter},
    isbn = {9780199570430},
    title = "{Occasionalism}",
    chapter = {9},
    pages = {64-78},
    booktitle = "{Causation and Laws of Nature in Early Modern Philosophy}",
    publisher = {Oxford University Press},
    year = {2009},
    month = {09},
    doi = {10.1093/acprof:oso/9780199570430.003.0010},
    eprint = {https://academic.oup.com/book/0/chapter/290816460/chapter-ag-pdf/44521609/book\_34299\_section\_290816460.ag.pdf},
}

@article{Curiel2020,
year = {2020},
title = {Schematizing the Observer and the Epistemic Content of Theories},
copyright = {http://arxiv.org/licenses/nonexclusive-distrib/1.0},
language = {eng},
author = {Curiel, Erik},
journal={arXiv preprint arXiv:1903.02182},
}

@article{Rynasiewicz,
 ISSN = {00318248, 1539767X},
 author = {Robert Rynasiewicz},
 journal = {Philosophy of Science},
 number = {4},
 pages = {572--589},
 publisher = {[The University of Chicago Press, Philosophy of Science Association]},
 title = {Rings, Holes and Substantivalism: On the Program of {L}eibniz Algebras},
 urldate = {2022-09-20},
 volume = {59},
 year = {1992}
}

@article{IntroISE,
    title={{\GG{1}} {I}ntroducing the {ISE M}ethodology: A Powerful New Tool for Topological Redescription},
    author={Daniel Grimmer},
    year={2023},
   journal={arXiv preprint arXiv:2303.04130},
}

@article{ISEEquiv,
      title={{\GG{2}} {I}n Search of New Spacetimes: Topological Redescription and the {ISE}-Equivalence Theorem}, 
      author={Daniel Grimmer},
      year={2023},
      journal={arXiv preprint arXiv:2306.08110},
}

@incollection{ExplainUnify,
	editor = {Philip Kitcher and Wesley Salmon},
	year = {1989},
	title = {Explanatory Unification and the Causal Structure of the World},
	author = {Philip Kitcher},
	booktitle = {Scientific Explanation},
	publisher = {University of Minnesota Press},
	pages = {410--505}
}

@incollection{KantDissertation,
place={Cambridge},
title={On The Form And Principles Of The Sensible And The Intelligible World [Inaugural Dissertation] (1770)},
booktitle={Theoretical Philosophy, 1755–1770},
publisher={Cambridge University Press},
author={Kant, Immanuel},
editor={Walford, David and Meerbote, Ralf},
year={1992}, 
pages={373–416}, 
collection={The Cambridge Edition of the Works of Immanuel Kant} 
}

@article{STepistemology,
author = {Dewar, Neil and Linnemann, Niels and Read, James},
title = {The epistemology of spacetime},
journal = {Philosophy Compass},
volume = {17},
number = {4},
pages = {e12821},
doi = {https://doi.org/10.1111/phc3.12821},
eprint = {https://compass.onlinelibrary.wiley.com/doi/pdf/10.1111/phc3.12821},
year = {2022}
}

@book{Poincare,
publisher = {Bloomsbury Publishing},
booktitle = {Science and Hypothesis},
isbn = {9781350026780},
year = {2018},
title = {Science and Hypothesis: The Complete Text},
language = {eng},
author = {Poincaré, Henri},
}

@article{pittphilsci8895,
title = {The nontriviality of trivial general covariance: How electrons restrict ‘time’ coordinates, spinors (almost) fit into tensor calculus, and 7/16 of a tetrad is surplus structure},
journal = {Studies in History and Philosophy of Science Part B: Studies in History and Philosophy of Modern Physics},
volume = {43},
number = {1},
pages = {1-24},
year = {2012},
issn = {1355-2198},
doi = {https://doi.org/10.1016/j.shpsb.2011.11.001},
author = {Brian Pitts},

}

@article{Acuna2016,
title = {Minkowski spacetime and {L}orentz invariance: The cart and the horse or two sides of a single coin?},
journal = {Studies in History and Philosophy of Science Part B: Studies in History and Philosophy of Modern Physics},
volume = {55},
pages = {1-12},
year = {2016},
author = {Pablo Acuña},
}

@article{linnemann2021constructivists,
      title={The Constructivist's Programme and the Problem of Pregeometry}, 
      author={Niels Linnemann and Kian Salimkhani},
      year={2021},
      journal={arXiv preprint arXiv:2112.09265},
}

@article{IVANOVA2015114,
title = {Conventionalism, structuralism and neo-{K}antianism in {P}oincaré's philosophy of science},
journal = {Studies in History and Philosophy of Science Part B: Studies in History and Philosophy of Modern Physics},
volume = {52},
pages = {114-122},
year = {2015},
issn = {1355-2198},
doi = {https://doi.org/10.1016/j.shpsb.2015.03.003},
author = {Milena Ivanova}
}

@book{read_2023,
place={Cambridge},
series={Elements in the Philosophy of Physics},
title={Special Relativity},
DOI={10.1017/9781009300599},
publisher={Cambridge University Press},
author={Read, James},
year={2023},
collection={Elements in the Philosophy of Physics}}

@incollection{Halvorson2016,
	author = {Hans Halvorson},
	booktitle = {The Oxford Handbook of Philosophy of Science},
	year = {2016},
	title = {Scientific Theories},
	pages = {585--608},
        chapter = {28},
	editor = {Paul Humphreys},
	publisher = {Oxford University Press}
}

@article{pittphilsci19728,
	author = {David Wallace},
	doi = {10.1016/j.shpsa.2022.01.015},
	journal = {Studies in History and Philosophy of Science Part A},
	number = {C},
	pages = {239--248},
	title = {Isolated Systems and Their Symmetries, Part {I}: General Framework and Particle-Mechanics Examples},
	volume = {92},
	year = {2022}
}

@book{PlatoMeno,
booktitle = {Meno},
isbn = {9781603847872 (electronic bk.)},
year = {1980},
title = {Meno},
language = {eng},
author = {Plato},
editor = {Grube, G. M. A},
publisher = {Hackett Publishing Company},
}

@book{Plato1997Cwer,
booktitle = {Complete works},
isbn = {9781603846707},
year = {1997},
title = {Complete works},
language = {eng},
author = {Plato},
editor = {Cooper, John M and Hutchinson, D. S.},
publisher = {Hackett Publishing Company},
}

\end{document}